\documentclass[apj]{emulateapj}
\usepackage[T1,T2A]{fontenc}
\usepackage[utf8]{inputenc}
\usepackage[russian,english]{babel}
\usepackage{enumitem}
\usepackage{url}
\usepackage{verbatim}
\usepackage{csquotes}
\def\kw{{Konus-\textit{Wind}}}
\def\rh{{\textit{RHESSI}}}

\usepackage{ulem,color,epstopdf,float} 
\usepackage{amsmath, graphicx}

\usepackage{hyphenat}
\newcommand{\al}{\bf \color{red}}

\received{March 25, 2019}
\revised{April 11, 2019}
\accepted{\today}

\begin{document}

\title{Gamma-ray emission from the impulsive phase of the 2017 September 06 X9.3 flare}
\author{Аlexandra L. Lysenko}
\affiliation{Ioffe Institute, Polytekhnicheskaya, 26, St. Petersburg, 194021 -  Russian Federation}
\email{alexandra.lysenko@mail.ioffe.ru}

\author{Sergey A. Anfinogentov}
\affiliation{Institute of Solar-Terrestrial Physics  (ISZF), Lermontov st., 126a, Irkutsk, 664033 -  Russian Federation}

\author{Dmitry S. Svinkin}
\affiliation{Ioffe Institute, Polytekhnicheskaya, 26, St. Petersburg, 194021 -  Russian Federation}

\author{Dmitry D. Frederiks}
\affiliation{Ioffe Institute, Polytekhnicheskaya, 26, St. Petersburg, 194021 -  Russian Federation}

\author{Gregory D. Fleishman}
\affiliation{New Jersey Institute of Technology, University Heights, Newark, NJ 07102-1982 -  United States}
\affiliation{Ioffe Institute, Polytekhnicheskaya, 26, St. Petersburg, 194021 -  Russian Federation}

\begin{abstract}
We report hard X-ray and gamma-ray observations of the impulsive phase of the SOL2017-09-06T11:55 X9.3 solar flare.
We focus on a high-energy part of the spectrum, $>100$~keV, and perform time resolved spectral analysis for a portion of the impulsive phase, recorded by the \kw\ experiment, that displayed prominent gamma-ray emission.
Given a variety of possible emission components contributing to the gamma-ray emission, we employ a Bayesian inference to build the most probable fitting model.
The analysis confidently revealed contributions from nuclear deexcitation lines, electron-positron annihilation line at 511\,keV, and a neutron capture line at 2.223\,MeV along with two components of the bremsstrahlung continuum.
The revealed time evolution of the spectral components is particularly interesting.
The low-energy bremsstrahlung continuum shows a soft-hard-soft pattern typical for impulsive flares, while the high-energy one shows a persistent hardening at the course of the flare.
The neutron capture line emission shows an unusually short time delay relative to the nuclear deexcitation line component, which implies that the production of neutrons was significantly reduced soon after the event onset.
This in turn may imply a prominent softening of the accelerated proton spectrum at the course of the flare, similar to the observed softening of the low-energy component of the accelerated electrons responsible for the low-energy bremsstrahlung continuum.
We discuss possible physical scenarios, which might result in the obtained relationships between these gamma-ray components.

\end{abstract}

\section{\label{sec_intro} Introduction}

A relatively quiet minimum of solar cycle \#~24  was interrupted by a series of strong flares that occured in September, 2017, including four X-class flares and 27 M-class flares.
The flare-productive active region (AR), AR~12673, had one of the largest magnetic flux emergence \citep{Sun2017} and  one of the strongest photospheric magnetic field of the order of 5,500~G \citep{Wang2018} ever observed.
An X9.3-class solar flare that occurred on September, 6, 2017, became the strongest flare of solar cycle \#~24. It produced numerous helioseismic waves, the so-called ``sunquakes'' \citep{Sharykin2018}.
In addition, this X9.3 flare on September, 6, and the X8.2 flare on September, 10, are of particular interest, because they were followed by a sustained gamma-ray emission observed by the Fermi-LAT \citep{Atwood2009} instrument during more than 10 hours at energies $>$100 MeV \citep{Longo2017a, Omodei2018}.
Multiple flux-rope eruptions accompanied both of these flares \citep[see, e. g.,][]{Hou2018, WangLiu2018}.
Also solar Energetic Particles (SEP) events associated with these flares were registered \citep{Nolfo2018}.

While emission from the impulsive phase of the 2017-Sep-10 X8.2 flare was well observed by different instruments in X-ray and microwave ranges \citep[see, e.g.,][and references therein]{Gary2018}, observations of the 2017-Sep-06 X9.3 flare impulsive phase are rather poor:  it occurred during ``nights'' of both \rh\ and \textit{Fermi} spacecrafts and did not have microwave coverage by any of available imaging instruments.

Here we report on the only available high-energy diagnostics of the impulsive phase of the 2017-Sep-06 flare in hard X-ray (HXR) and gamma-ray ranges at photon energies up to $\sim$15\,MeV obtained with the \kw\ instrument.
Gamma-ray emission from this flare is the primary focus of this work.

In contrast to the HXR emission from solar flares, which is produced by a single emission process, the bremsstrahlung, a variety of distinct emission processes contributes at the gamma-ray domain.
The bremsstrahlung from relativistic electrons and positrons still produces the flare gamma-ray continuum, while accelerated ions could manifest themselves through more spectral components via numerous nuclear reactions.
These nuclear reactions result in gamma-ray emission from nuclear deexcitation lines, the line of electron and positron annihilation at 511\,keV, line complex from fusion reactions between $\alpha$-particles, the line from neutron capture by a proton at 2.223\,MeV, and continuum from the pion decay \citep[see, e.g.,][and references therein]{Ramaty1975, Vilmer2011, Ackermann2012}.

Accelerated ions interact with the ambient plasma through inelastic scattering, spallation or fusion reactions.
The resulting nuclei are in excited states and their transition to the ground state is accompanied by the emission of characteristic gamma-quanta \citep{Ramaty1975, Murphy2007, Murphy2009}.
When accelerated protons or $\alpha$-particles interact with the heavier ions of the ambient plasma, narrow nuclear gamma-ray lines (FWHM$\sim$2\,\%) are emitted due to relatively low  thermal velocities of the plasma ions.
But interaction of heavier accelerated ions with ambient protons and $\alpha$-particles result in emission of broad (FWHM$\sim$20\,\%) nuclear gamma-ray lines.
Thus, the ratio between fluxes in narrow and broad lines allows estimating abundances of accelerated ions and ions in ambient plasma \citep{Murphy2007}. On top of that, some gamma-ray lines permit diagnostics of the $^3$He abundance \citep{Murphy2016}, which is important given that the acceleration in solar flares can result in strong enrichment of the $^3$He ions.

Nuclear collisions also produce free neutrons. A fraction of those neutrons escape the Sun, while others reach the photosphere, where they can be captured by protons with the formation of deuterium.
This reaction results in the emission of a very narrow (FWHM<0.1\,keV) line at 2223\,keV \citep{Shih2009}.

The presence of the pion decay emission in a flare spectrum evidences, that protons were accelerated up to $\sim$300\,MeV energies, which is the threshold for pion production reactions \citep{Murphy1987}.
The decay of neutral pions $\pi^0$ gives two $\gamma$-quanta with energies 67.6\,MeV in the pion rest frame, thus observed emission from the $\pi^0$ decay is highly Doppler-shifted and represents a broad peak at $\sim$70\,MeV with FWHM of about one order of magnitude~\citep{Crannell1979}.

Charged pions $\pi^+$ and $\pi^-$ decay to ultrarelativistic positrons and electrons. Some positrons interact with dense chromo/photospheric plasma of the flaring loop footpoints to produce gamma-ray continuum via the bremsstrahlung at energies $\geq$10\,MeV and to eventually annihilate with electrons giving prompt annihilation line at 511\,keV \citep{Crannell1979, Murphy1987}. In addition, nonthermal positrons can be produced by unstable $\beta^+$-active nuclei, which also contribute to the gamma-ray continuum and delayed annihilation gamma-ray emission \citep{Kozlovsky1987}.

Although there are many nuclear processes, which, when detected, can trace acceleration of ions, some flares confidently detected in gamma-ray range do not show any signature of the gamma-ray lines above the bremsstrahlung continuum. Such flares have come to be known as ``electron-dominated'' ones \citep{Marschhauser1994, Trottet1998, Rieger1998}. This implies that the number of distinct spectral components to be included in the fit function cannot be taken ``for granted,'' while has itself to be determined from the data analysis.

In summary, gamma-ray emission of solar flares offers a unique tool of studying ion acceleration and transport in flares, which is needed for detailed understanding of how the solar flare works given that the accelerated ions can make a significant contribution to the total flare energetics \citep{Emslie2004, Emslie2012}.
In this study, we focus on a spectral analysis of the impulsive phase of the 2017-Sep-06 X9.3 flare to define the actual spectral components, quantify their spectral evolution, and estimate energetics of the accelerated ions.

\section{\label{sec_obs} Observations}

\kw\ instrument is the X-ray/gamma-ray spectrometer employed for studies of gamma-ray bursts and solar flares \citep{Aptekar1995, Palshin2014}.
\kw\ works in two modes: the waiting mode and the triggered mode.
In the waiting mode, the count rates in three wide energy bands G1~($\sim$20--80\,keV), G2~($\sim$80--300\,keV) and G3~($\sim$300-1200\,keV) are measured with 2.944 s time resolution.
When the count rate in the G2 band exceeds a $\sim$9$\sigma$ threshold above the background on one of two fixed timescales, 1\,s or 140\,ms, the instrument switches into the triggered mode.
In the triggered mode\footnote{Data for all flares registered by \kw\ in the triggered mode are available online in $KW$-$Sun$ database via \url{http://www.ioffe.ru/LEA/kwsun/} and \url{http://www.ioffe.ru/LEA/Solar/}}, the count rates in the three energy bands are recorded with time resolution varying from 2\,ms up to 256\,ms. These light curves, with a total duration of $\sim$230 s, also include 0.512\,s of pre-trigger data.
Spectral measurements are carried out starting from the trigger time $t_0$, in two overlapping energy intervals, with 64 spectra being recorded for each interval over a 63-channel, pseudo-logarithmic energy scale.
Accumulation time for the first four spectra is fixed at 64\,ms and for the last eight spectra -- at 8.192\,s; for the remaining spectra, the accumulation time varies from 256\,ms to 8.192\,s according to the count rate in G2 channel: for time intervals with higher intensity, the accumulation times are shorter.
After the end of the trigger record, \kw\ is inactive for $\sim$1\,hour because of the data readout, and only the light curve in G2 channel with $\sim$3\,s temporal resolution is available for this hour.
According to uninterrupted observations in the hard G2 channel the flare started $\sim$70\,s before the trigger time and ended $\sim$590\,s after; thus, the impulsive phase of the flare lasted for $\sim$11\,minutes.

\begin{figure*}\centering
\includegraphics[scale=0.6]{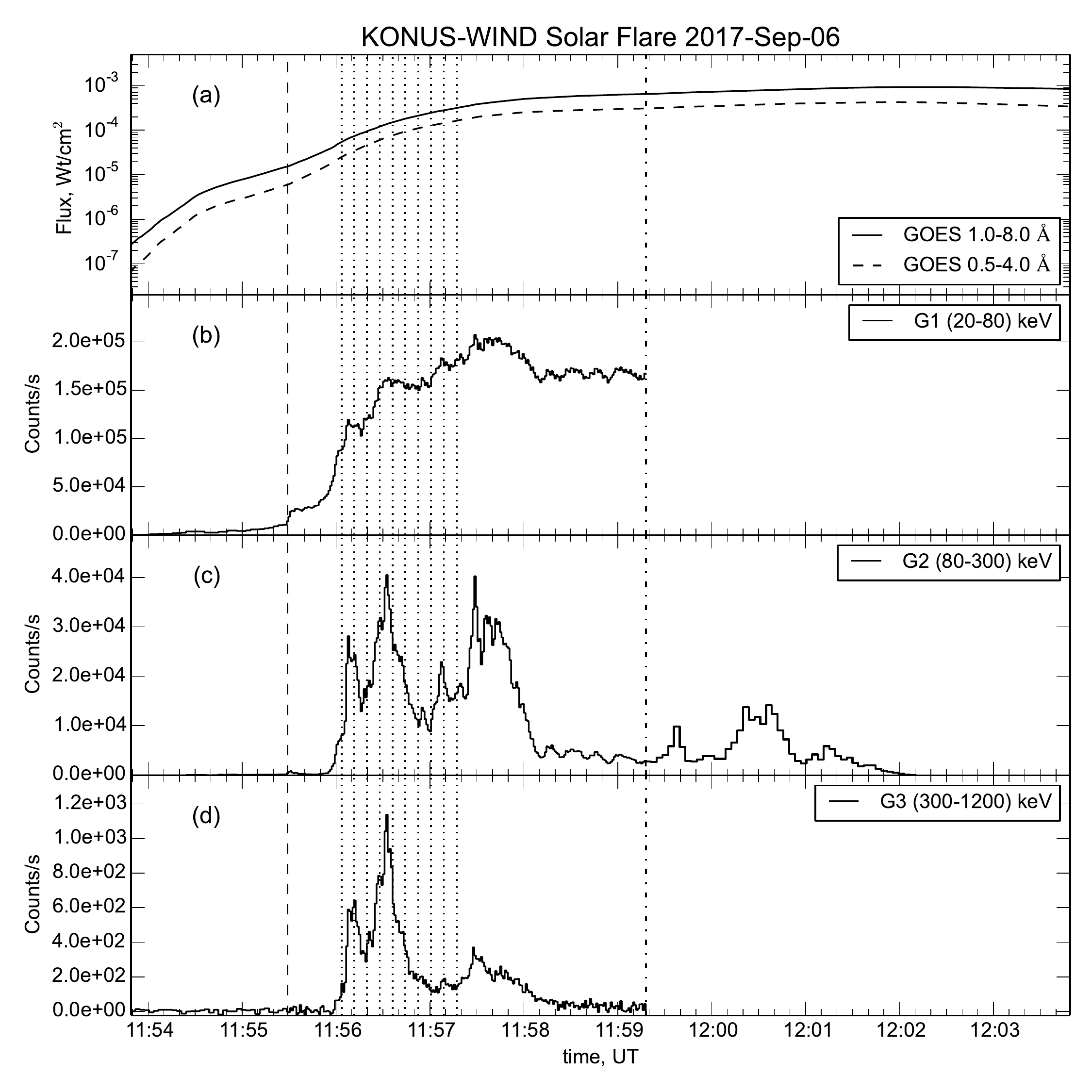}
\caption{\label{fig_lc} Time profiles of 2017-Sep-06 X9.3 flare in X-ray range: (a) time profiles of GOES 1--8\,\AA\ and 0.5--4.0\,\AA\ channels; (b), (c), (d): \kw\ time profiles in G1, G2 and G3 channels correspondingly. Dashed vertical line indicates \kw\ trigger time, dotted vertical lines mark the multichannel spectra accumulation intervals used for fitting, dash-dotted line denotes the end of the high temporal resolution record in G1, G2 and G3 channels. }
\end{figure*}

\kw\ observed 2017-Sep-06 X9.3 flare in the triggered mode since 11:55:29.0~UT\footnote{Hereafter, the time is corrected for the light propagation from \kw\ to the Earth center for an easier compatibility with the near-earth spacecrafts.
Selection of the center of the Earth gives uncertainties within 20\,ms for the near-earth and ground-based instruments, thus all time stamps are rounded to tenth of a second.}.
Light curves in the G1, G2 and G3 energy bands are presented in Figure~\ref{fig_lc}. Recording of the light curves with high time resolution lasted till 11:59:18.7 (Figure~\ref{fig_lc}).
During this time, almost monotonic count rate increase was observed in the G1 channel, while a few peaks can be distinguished in the G2 channel (the main peaks are near 11:56:00~UT, 11:56:30~UT, and 11:57:20~UT), and two strongly pronounced peaks that occurred in the hard G3 channel near 11:56:00~UT and 11:56:30~UT were followed by a weak blurred peak near 11:57:30~UT.
After the end of the trigger record, 3 more maxima occurred in the G2 channel, but, unfortunately, there are no high-resolution measurements for this period, which would facilitate a detailed spectral analysis.

For this flare, the multichannel energy spectra in two partially overlapping energy bands are available for the interval from 11:55:29.0 to 11:57:17.1, which covers the first two maxima in the G2 and G3 bands.
The first band (PHA1) extends from 21\,keV to 1225\,keV and the second band (PHA2) corresponds to energies from 400\,keV to 15.4\,MeV.

Spectral analysis relies on energy calibration of the data.
The energy calibration for multichannel data in $KW$-$Sun$ database is performed using an automated procedure, which is sufficiently accurate for the continuum component of the spectrum, but insufficient for analysis of the gamma-ray lines; thus, we recalibrated the multichannel data manually to a higher accuracy.

For the context, we  employed the GOES time profiles in the 1--8\,\AA\ and 0.4--4.0\,\AA\ channels, which are plotted in Figure~\ref{fig_lc}(a). The \kw\ observations, Figure~\ref{fig_lc}(b)--(d),  cover the rise phase of the thermal soft X-ray (SXR) emission and the beginning of the broad main peak.

\section{\label{sec_analysis} Data Analysis}

\subsection{\label{sec_an_mult} Multichannel data preparation}

As the main subject of this work is gamma-ray emission, we fitted only those multichannel spectra which demonstrate an excess over the background above $\sim$500\,keV, i.e. since 11:56:03.8~UT till the end of multichannel spectra accumulation (11:57:17.1~UT).
For the time period from 11:56:03.8~UT to 11:56:11.7~UT, spectra are available with the accumulation time of 256\,ms; however, such a short exposure gives insufficient number of  counts at higher energies, so we summed these spectra up to form one spectrum with the accumulation time of 7.936\,s, which is close to the accumulation times of the remaining eight spectra, 8.192\,s.
Thus, we analyzed multichannel spectra for nine time intervals, which are listed in Table~\ref{table_best_model} and are shown in Figure~\ref{fig_lc}, along with the time-integrated spectrum for the entire time range of 11:56:03.8~UT to 11:57:17.1~UT.

The spectrum accumulation ended before the end of the flare; therefore, we were forced to take a background spectrum from another event, registered by \kw\ in the same day in the same detector in the triggered mode.
The differences between background rates for this event and the flare, estimated according to G1, G2 and G3 channels, are within 2.3\,\%, 2.0\,\% and 1.0\,\% correspondingly, which were summed in quadrature and considered as a $\sim$3\,\% systematic uncertainty in the background \citep{LysenkoKwsun}.


For such an intense flare, at very high count rates, up to $\sim$10$^5$\,cts/s in the soft PHA1 band, the spectrum shapes are distorted due to a pulse-pileup effect, which is not accounted for by a standard \kw\ dead-time correction procedure (i.e., a simple non-paralyzable dead-time correction in each of the measurement bands).
To minimize the spectrum deformations, the PHA1 counts were corrected using the rate- and spectral-model-dependent coefficients obtained for each spectral channel from Monte-Carlo modeling of the pulse pile-up effect (for more detailed description of the pulse-pileup corrections see Appendix).
After that, to take into account other instrumental effects which are non-negligible at the high count rates, such as the differential nonlinearity of the instrument's analog-to-digital converters, a systematic error of $\sim$4\,\% has been added in quadrature to the statistical uncertainty in the count spectra.

For all nine time intervals, the spectral fitting of the first energy range in \verb"XSPEC"~\verb"12.9.0"  \citep{Arnaud1996} revealed an energy break-down at about 100\,keV, which cannot be explained by the pile-up effect only.
Spectral flattening at these energies could be associated with either a break in the energy spectrum of accelerated electrons or with some other physical effects \citep{Lysenko2018}.
As the focus of this work is the higher energy emission, we chose the energy of this break, 100\,keV, as the lower bound for the spectral analysis, while we took the end of the second energy range, 15.4\,MeV, as the high-energy bound.

\subsection{\label{sec_an_mult_new} Spectral components for fitting}

Selection of specific shapes of the spectral components to be included in the fit requires a special care.
The bremsstrahlung continuum from accelerated electrons can  in some cases be described by a simple power-law model, PL \citep{Share2000, Lin2003}.
Sometimes, however, a spectral flattening is observed at higher energies. Often, bremsstrahlung continuum can be well described by a broken power law, BPL \citep[see, e.g.,][]{Share2003}:

\begin{equation}
\label{eq_2PL}
I(E) \propto \begin{cases} E^{-\gamma_1} & E \le E_{\mathrm{break}} \\
 E^{-\gamma_2} & E > E_{\mathrm{break}} 
\end{cases}
\end{equation}
In other cases, a cutoff at higher energies is observed, so that the continuum can be represented by a sum of a PL component and a flatter power law with an exponential cutoff, CPL \citep[see, e.g.,][]{Marschhauser1994, Petrosian1994, Ackermann2012}:

\begin{equation} \label{eq_cpl}
I(E) \propto E^{-\gamma}\exp \left(-\frac{E}{E_{\mathrm{cut}}} \right)
\end{equation}

In this work, we also consider a broken power law with an exponential cut-off (BPLexp) which is  a combination of (\ref{eq_2PL}) and (\ref{eq_cpl})
\begin{equation}\label{eq:bpl_exp}
 I( E) \propto \begin{cases}
	E^{-\gamma _{1}}  & E \le E_{\mathrm{break}}\\
	E^{-\gamma _{2}}\exp\left( -\frac{E}{E_{_{\mathrm{cut}}}}\right)  & E > E_{\mathrm{break}}
	\end{cases}
\end{equation}

Some of the nuclear deexcitation lines (e. g., 6.129\,MeV from $^{16}$O, 4.438\,MeV from $^{12}$C and some others) are strong enough to stand out cleanly against the continuum, while many weaker lines from close nuclear levels merge into a continuum and individual fitting of these lines is not possible.
To facilitate the spectral fitting of a whole variety of the gamma-ray lines, \cite{Murphy2009} developed spectrum templates for nuclear deexcitation lines based on common conditions in the solar atmosphere for both narrow and broad lines (see Introduction) and these templates were added to the \verb"OSPEX" package \citep{Schwartz2002}.
Unfortunately, the \kw\ spectral resolution does not permit separating the narrow and broad lines; thus, we had to use a combined template also available from the \verb"OSPEX" package, NUCLEAR, where narrow and broad lines are mixed for averaged abundances observed in 19 SMM flares \citep{Murphy2009}.

Individual gaussian profiles were added for the annihilation and neutron capture lines. The line of the e$^+$--e$^-$ annihilation was modeled as a gaussian with maximum fixed at 511\,keV and width fixed at $\sigma$=3\,keV \citep{Share2003}.
The neutron capture line was modeled as a gaussian with the peak at 2223\,keV and $\sigma$ fixed at 0.1\,keV \citep{Shih2009}.
We do not add any gamma-ray emission from the pion decay here as it is expected at energies about tens of MeV which is outside the \kw\ energy range.

\subsection{\label{sec_bayes} Bayesian Inference}

As has been said in the Introduction, the first needed step is to identify what spectral components are present in the observed gamma-ray emission. This task is not as simple as it might look especially for the moderate-resolution data. Indeed,
the least squares fitting is not generally capable of favoring the presence/absence of a given component \citep{Protassov2002}.
In some cases, the least square method can be used as an $F$-test \citep{Bevington1969}, which may favor a model based on a decrease of the $\chi^2$ metrics relative to the associated reduction of the number of the degrees of freedom.
This approach, however, can only by conclusively used if: (i) the competing models are ``nested'' -- one model can be retrieved from the other by adding more parameters, e.\,g., a simple power law and  a broken power law; and (ii) none of the parameters is close to its bounds \citep{Protassov2002}.
These conditions are obviously not fulfilled in our case.

A meaningful model comparison should explore the entire parameter space to address the following questions:
\begin{itemize}
	\item How close is the best fit to the actual observations?
	\item Do the models have sufficient number of parameters (without over-fitting)?
	\item What model is better confined in the parameter space (has narrower uncertainties)?
	\item What model is better confined in the observational data space (predicts possible observations closer to the actual data)?
\end{itemize}

An adequate quantitative comparison of competing models that accounts for all these points can be performed within a framework of a Bayesian Inference by calculating a so-called \textit{Bayes Factor}.
Due to its universality and robustness, the Bayesian analysis  has recently become a gold standard in different kinds of data processing problems including HXR and gamma-ray spectral analysis \citep[][and others]{vanDyk2001, Trotta2008, Abdo2009, Buchner2014}.

The Bayesian inference implies the investigation of a Posterior probability distribution function (PDF) $P(\theta|D,M)$ which
denotes the probability that the model parameters are equal to $\theta$ given that the observational data $D$ are produced with a model $M$.
	The posterior PDF can be calculated using the \textit{Bayes theorem}:
\begin{equation} \label{eq:Bayes theorem}
P(\theta|D,M) = \frac{P(D|\theta,M)P(\theta)}{P(D|M)},
\end{equation}
where $\theta = \theta_i$ is the set of parameters for the model $M$, $P(\theta)$ is their prior distribution, based on our a priori knowledge about these parameters, and $P(D|\theta,M)$ is the likelihood function or the conditional PDF of  measuring observational data $D$ supposing that they are generated by the model $M$ with parameters  $\theta$.

In this study, we assume that the measurement errors are normally distributed. This gives the following likelihood function:

\begin{equation}
P(D|\theta,M) = \frac{1}{(2\pi \sigma_i^2)^{\frac{N}{2}}}\prod_{i=1}^{N} \exp \left\lbrace  -\frac{[C_i - C_M(E_i,\theta)]^2}{2\sigma_i^2}\right\rbrace.
\label{eq:mcmc_fit likelihood}
\end{equation}
Here, $N$ is the number of energy channels, $E_i$ is the central energy of $i$th channel, $C_i$ is the observed number of counts detected in $i$th channel, $\sigma_i$ are known measurement errors, and  $C_M(E_i,\theta)$ denotes the model prediction of count number in channel $i$ depending upon the model parameters $\theta$.

The normalization constant $P(D|M)$ in the denominator of Eq.~\ref{eq:Bayes theorem} is the \textit{Bayesian Evidence} or marginalised likelihood

\begin{equation} \label{eq:Evidence}
Z=P(D|M)=\int P(D|\theta,M)P(\theta) d\theta.
\end{equation}

Two models $M_1$ and $M_2$  can be quantitatively compared by calculating the \textit{Bayes Factor}
\begin{equation} \label{eq:Bayes factor}
B_{12} = \frac{P(D|M_1)}{P(D|M_2)}.
\end{equation}
Values  of $B_{12}$ greater  than  3,  20  and  150  are interpreted as \lq\lq positive\rq\rq , \lq\lq strong\rq\rq, and \lq\lq very strong\rq\rq\  evidence for model $M_1$ over model $M_2$, respectively.

If one have several competing models $M_i$ with evidences $Z_i$, the posterior probability $P_i$ of $i$-th model can be calculated using the Bayes theorem:
\begin{equation}\label{eq:posterior_model_prob}
P_i = P(M_i|D) = \frac{P(D|M_i)P(M_i)}{P(D)},
\end{equation}
where $P(M_i)$ is the prior probability for a model $M_i$, and $P(D)$ is the normalization  constant defined as
$$
P(D) = \sum\limits_{j=1}^{N_M} P(D|M_j)P(M_j),
$$
where $N_M$ is the total number of the competing models.
In this work, we assume that our list of models is complete and all models are a priori equally probable (we will return to this assumption below); therefore,

\begin{equation}\label{eq:model_prob}
P_i = \frac{P(D|M_i)}{\sum\limits_{j=1}^{N_M} P(D|M_j)} = \frac{Z_i}{\sum\limits_{j=1}^{N_M} Z_j}.
\end{equation}

By calculating posterior probabilities $P_i$ for the case of two competing models, we can establish correspondence between the Bayes factor $B_{12}$ and the posterior probabilities $P_i$.
This approach gives us that  \lq\lq positive\rq\rq , \lq\lq strong\rq\rq, and \lq\lq very strong\rq\rq\  evidence for model $M_1$ over all other models corresponds to $P_1 = 0.75$, $P_1\approx 0.95$, and   $P_1 \approx 0.99$, respectively.

For drawing samples from the  posterior probability distribution (\ref{eq:Bayes theorem}), we use our own implementation of
the Metropolis-Hastings algorithm for Markov Chain Monte-Carlo (MCMC) sampling.
The evidence integrals defined by Eq. (\ref{eq:Evidence}) needed for the Bayes factor calculation are computed using the Monte-Carlo integration with importance sampling.
More detailed description of our MCMC code\footnote{The MCMC code is available at \url{https://github.com/Sergey-Anfinogentov/SoBAT}}  can be found in \citet{2017A&A...600A..78P}
where it was successfully applied for the seismological inference of the transverse density profiles in oscillating coronal loops.

\subsection{\label{sec_fit_inegral} Selection of the fitting model}

We select the ultimate fitting model for our analysis in two steps: (1) selection of an analytical model for the continuum spectrum and (2) evaluation of evidence for other spectral components such as 511\,keV, 2.2\,MeV lines, and nuclear deexcitation lines. For the model selection we used the time-integrated spectrum (interval \#~0 in Table~\ref{table_best_model}).

To evaluate possible models for the continuum photon spectrum, we consider five options: a single power law (PL),  a sum of two power laws (PL + PL2), a sum of a power law and a power law with exponential cut-off (PL + CPL), a broken power law (BPL), and a broken power law with an exponential cutoff at higher energies (BPLexp; see Eq.~\ref{eq:bpl_exp}).
To account for the nuclear spectral components, all of them are included in  the analysis at this step.

\begin{figure*}\centering
  \includegraphics[width=1.\linewidth]{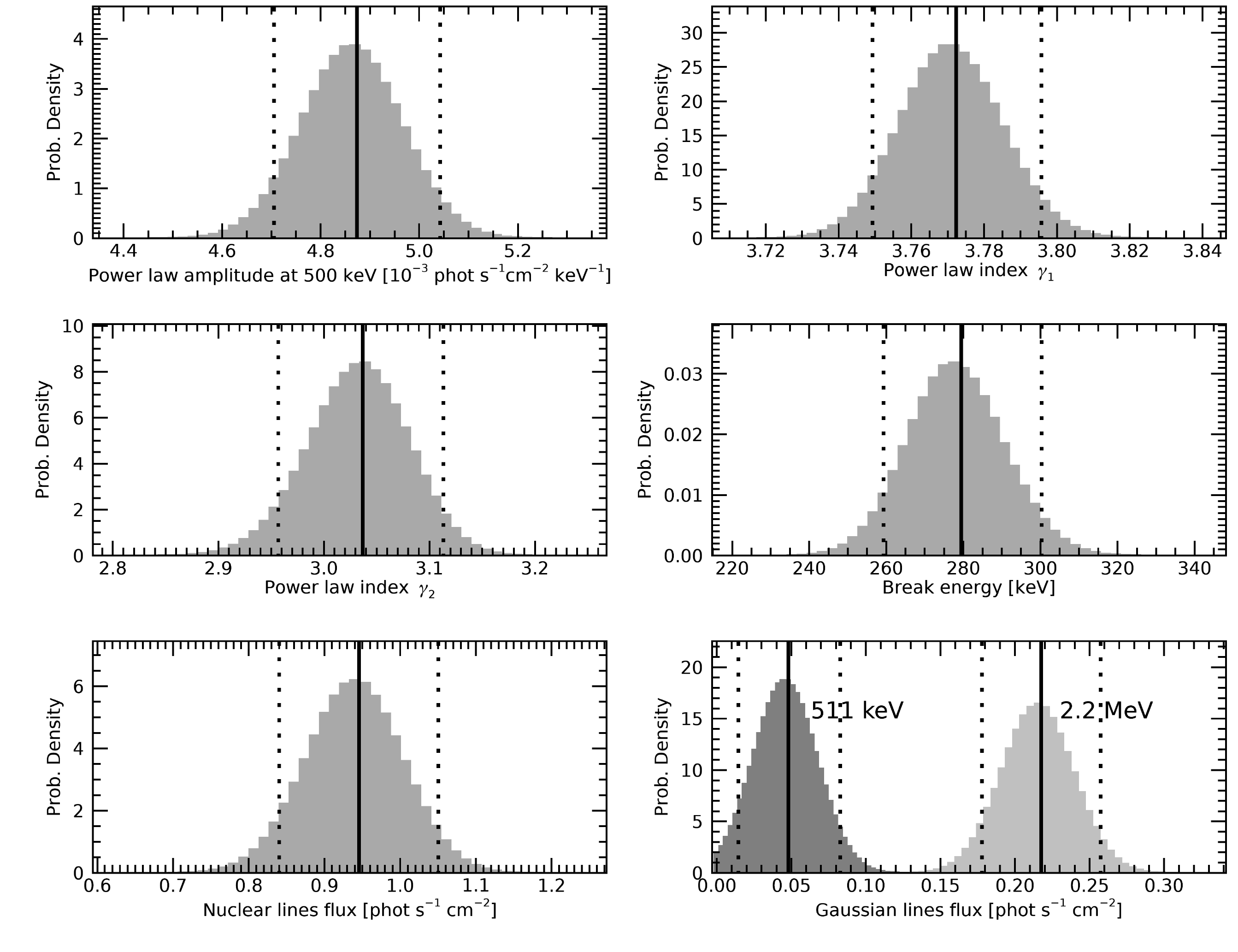}
  \caption{\label{fig_hist_bpl} Fitting parameter distributions for the time-integrated spectrum for BPL + 511\,keV~line + nuclear + 2.2\,MeV~line model. Vertical solid lines correspond to the most probable parameter values and the dotted  lines show 90\,\% credible intervals.}
\end{figure*}

The prior information has been set in the form of non-informative uniform prior distributions in linear scale with the following ranges: 1--10 for the power-law index in the PL component, 1--3 for the power-law index in the CPL component, 1--10 for the second power law index in BPL component, 0--1\,photons\,$\mathrm{cm}^{-2}\mathrm{s}^{-1}\mathrm{keV}^{-1}$ for the PL and CPL amplitudes at 500\,keV, 1--15\,MeV for the cutoff energy in the $CPL$ component and 0--10\,photons\,$\mathrm{cm}^{-1}\mathrm{s}^{-2}$ for the flux in the remaining components (NUCLEAR template for the nuclear deexcitation lines,  511\,keV and 2.2\,MeV lines).

For each model, we calculated Bayesian evidences (Eq. \ref{eq:Evidence})  and model posterior probabilities  (Eq. \ref{eq:model_prob}).
The  comparison of five alternative models is presented in Table~\ref{table_model_comp_global} for the time-integrated spectrum.
According to our calculations, the most probable model for the continuum part of the spectrum is the BPL model  with the posterior probability of 0.76.
The second significant model is BPLexp with the posterior probability of 0.24.
Other options  can be safely excluded since their total probability  is  less than $10^{-3}$.

Note that our first choice was the commonly employed PL+CPL model.
However, our Bayesian analysis reveals that PL+CPL model has a vanishing probability of $<10^{-3}$ and therefore, this model is much less applicable to this particular data set than BPL or BPLexp.
The main cause of this is that the models based on a simple sum of two power law components (PL+PL2, PL+CPL) imply a cross-talk between the components, meaning that one can obtain nearly the same models with different combination of PL, CPL and nuclear deexcitation lines components.
This cross-talk causes very high uncertainties for the normalization coefficients of the PL components.
The broken power law models (BPL and PBLexp) do not have this cross-talk and, therefore, provide a much better description of our data, which results in higher posterior probabilities.

For the subsequent analysis, we selected BPL model which has the highest posterior probability. The exponential cutoff, however, can exist, but the lack of counts at higher energies doesn't allow us to reliably define the cutoff energy $E_{cut}$. Based on analysis of time-integrated spectrum with BPLexp component we can say that $E_{cut}$ is greater than 5.8 MeV with the probability of 90\,\%.

To evaluate the need of each nuclear spectral component in fitting the observed spectra, we analyzed time-integrated spectrum with different combinations of these spectral components and the already validated broken power law model for the continuum spectrum.
For each model we calculated  evidences and model probabilities that are  listed in Table~\ref{table_model_comp_components}.
From eight possible combinations of components only two models  have probability above 1\%.
The most probable (94\%) model includes all components but 511\,keV line.
The probability of the model with all components is slightly less than 6\%.
Despite the  probability of the presence of the annihilation line  in the integrated spectrum is only 6\%, we decided to keep all components for the subsequent analysis, because the 511\,keV line can at times appear more cleanly,  while remaining less significant in the time-integrated spectrum. Indeed, having strong evidence for other nuclear components, we expect that the annihilation line is also likely present in the flare emission. From the viewpoint of the employed hear Bayesian statistics, this could have been accounted from the very beginning by giving a higher prior probability to the model containing all three nuclear components compared with any model that includes only a subset of them, because, if the nuclear de-excitation spectral component is present, the presence of other nuclear processes giving rise to other components is highly likely.

\begin{deluxetable}{llcc}
	\tablecolumns{4}
	\tablecaption{ \label{table_model_comp_global}Bayesian evidences and probabilities for different fitting models of continuum component for integrated spectrum.}
	\tablehead{\colhead{No.} & \colhead{Model} & \colhead{ln(Z)} & \colhead{Probability} }
	\startdata
	1 & BPL  & -173 & 0.76\\
	2 & BPLexp  & -174 & 0.24\\
	3 & PL &-340 & 0\tablenotemark{a}\\ 
	4 & PL + PL2 & -181 & 0\tablenotemark{a}\\
	5 & PL + CPL & -183 & 0\tablenotemark{a}\\
	\enddata
  \tablenotetext{a}{less than $10^{-3}$}
\end{deluxetable}

\begin{deluxetable}{lccccc}
	\tablecolumns{6}
	\tablecaption{ \label{table_model_comp_components}Bayesian evidences and probabilities for fitting models with different combinations of spectral components\tablenotemark{a}.}
	\tablehead{\colhead{No.} & \colhead{511 keV} & \colhead{2.2 MeV} &\colhead{nuclear} &  \colhead{ln(Z)} & \colhead{Probability} }
	\startdata
	1 &+ &+&+&-173& 0.056 \\
	2 & +&+&- &-256&0\tablenotemark{b}\\
	3 & +&-&+&-208& 0\tablenotemark{b}\\
	4 & +&- &-&-289&0\tablenotemark{b}\\
	5 &- &+&+&-170& 0.944\\
	6 &- &+& -&-252&0\tablenotemark{b}\\
	7 & -&-&+&-205& 0\tablenotemark{b}\\
	8 & -&- &-&-320& 0\tablenotemark{b}\\
	\enddata
	\tablenotetext{a}{Here the ``+'' sign means the presence of the corresponding component in the model, while the ``-'' sign means that the component is not added to the model.}
	\tablenotetext{b}{less than $10^{-15}$}
\end{deluxetable}

Thus, our analysis revealed the most appropriate model, which is a combination of gamma-ray lines on top of the broken power law dependence: \texttt{BPL + 511\,keV~line + nuclear + 2.2\,MeV~line}.
The marginal posterior probabilty distributions for the parameters of this model are presented in Figure \ref{fig_hist_bpl}.

\begin{figure}\centering
  \includegraphics[scale=0.55]{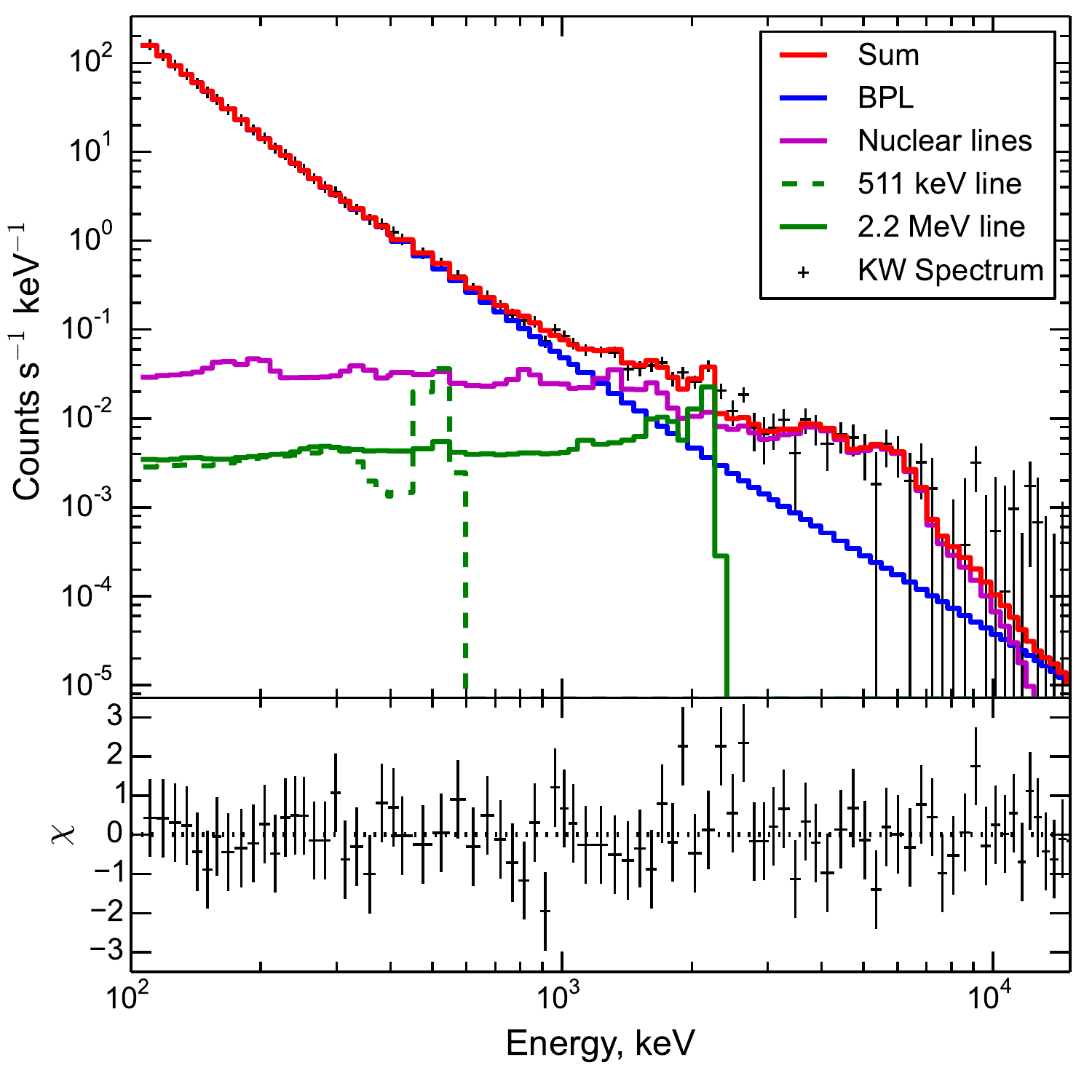}
  \caption{\label{fig_spec_ex} Fitting of the time-integrated spectrum by the BPL+NUCLEAR+511~keV+2.2~MeV model. The component color codes are given in the panel. The \kw\ data are shown using symbols, where the horizontal dashes indicate the energy range of each data point, while the vertical dashes indicate the corresponding statistical errors. The lower panel shows the fit residual.   }
\end{figure}

The time-integrated spectrum with the best-fitting model is presented in Figure~\ref{fig_spec_ex}, while the fitting parameters are listed in Table~\ref{table_best_model}, interval~\#~0.
The spectral break $E_{break}$ in the BPL component is found at $\sim$280\,keV with the power-law indices $\gamma_1$=3.78 below the break and $\gamma_2$=3.04 above the break, which are not much different.
To check if the emission in the continuum at lower and at higher energies is produced by different particle populations we added the BPL amplitude at 100\,keV, and that at 10\,MeV to the Table separately.
We also calculated residual $\chi^2$ between the observed spectrum and the model with parameters obtained by the Bayes inference, and based on these residuals and the number of degrees of freedom (dof), we estimated the probability for the fitting results (Prob.) and added it to the Table.

\subsection{\label{sec_fit_all}Analysis results for the time resolved spectra}

\begin{figure*}\centering
\includegraphics[scale=0.6]{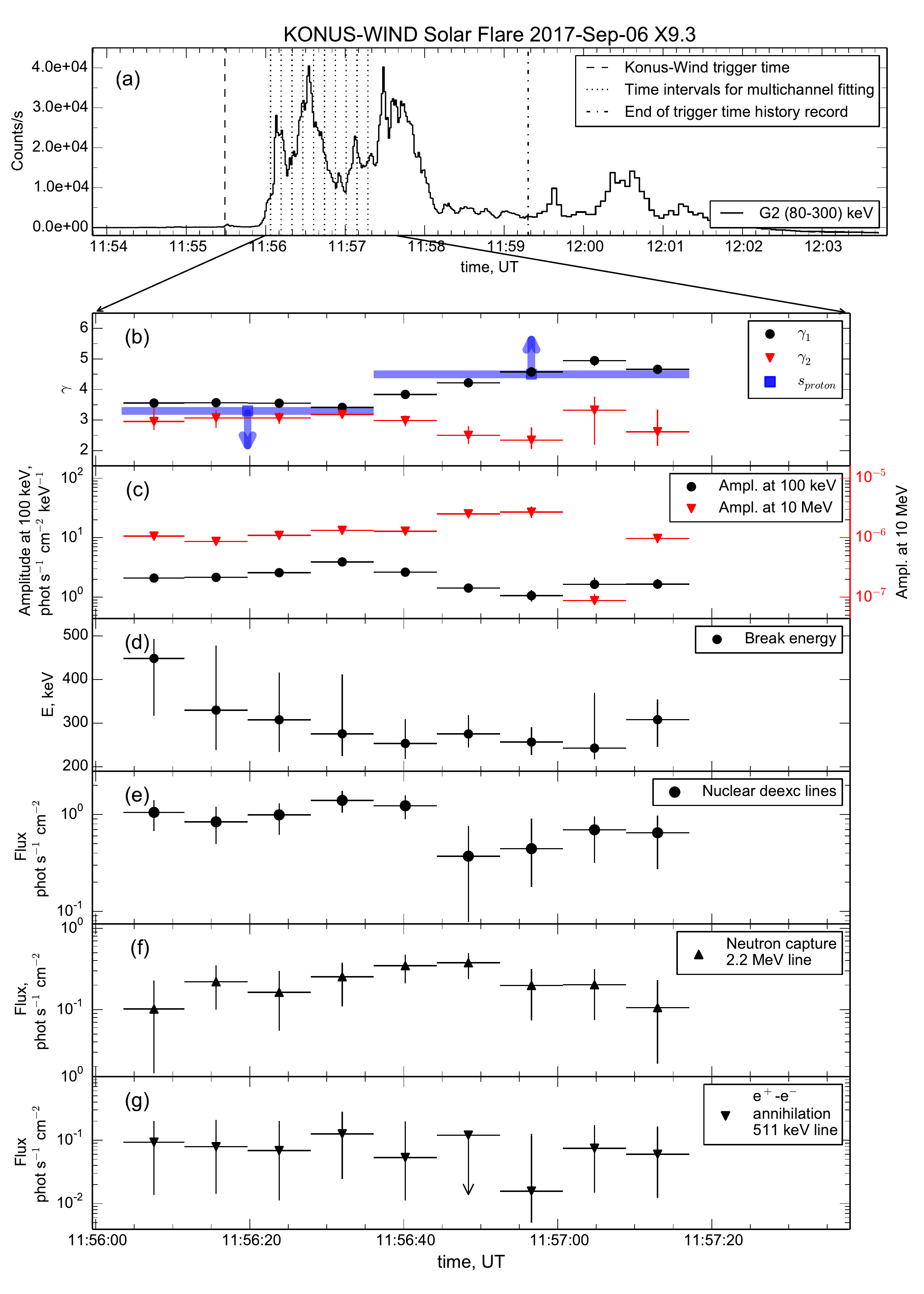}
\caption{\label{fig_mult} Spectral evolution of 2017-Sep-06 X9.3 flare in HXR and gamma-ray ranges obtained with \kw\ data in the context of the time profile of the \kw\ G2 channel. (a) The vertical dashed lines denote  time intervals used for studied multichannel spectra. (b) Time evolution of the photon power-law indices at the low energies, $\gamma_1$ (black circles), and at the higher energies, $\gamma_2$ (red triangles), along with the estimated limits for the proton power-law index $s$ (blue squares), see Section~\ref{sec_neut}. (c) Amplitudes for the broken power law component, BPL at 100\,keV (black circles) and at 10\,MeV (red triangles).  (d) The break energy $E_{break}$ for BPL. (e) The flux in the nuclear deexcitation lines; NUCLEAR template. (f) The flux in neutron capture line at 2.223\,MeV. (g) The flux in the electron-positron annihilation line at 511\,keV.
}
\end{figure*}

The technique described above was applied to the time resolved 9 intervals.
The results are listed in Table~\ref{table_best_model} and plotted in Figure~\ref{fig_mult}.
Amplitudes of the BPL component at 100\,keV and 10\,MeV evolve coherently before the main HXR peak at $\sim$11:56:35 with the power-law indices $\gamma_1$ and $\gamma_2$ not much different from each other and the spectral break spread between $\sim$250 and $\sim$ 500\,keV.
However, after the peak, the spectral break becomes more and more prominent with softening of the low-energy power-law component and hardening of the high-energy power-law component.
All time intervals (including the time-integrated spectrum) demonstrate significant contribution from the nuclear deexcitation and the neutron capture lines, but the confidence bands for the flux in the positron-electron annihilation line are only marginally higher than zero.

\clearpage
\begin{turnpage}
\begin{deluxetable}{lcccccccccccc}
  \tablecolumns{13}
  \tablecaption{ \label{table_best_model}Summary of Konus-Wind spectral fits in 100~keV--15~MeV energy range with $BPL+511\,keV$~$line+nuclear+2.2\,MeV$~$line$ model.}
    \tablehead{\colhead{No.} & \colhead{t$_{\rm start}$} & \colhead{t$_{\rm stop}$} & \colhead{BPL, $\gamma_1$}  & \colhead{BPL, $A_1$\tablenotemark{a}} & \colhead{$E_{break}$} & \colhead{BPL, $\gamma_2$}  & \colhead{$BPL$, $A_2$\tablenotemark{b}} & \colhead{$F_{511}$} & \colhead{$F_{nuclear}$} & \colhead{$F_{2.2}$} &  \colhead{$\chi^2$/dof} & \colhead{Prob.} \\
    \colhead{} & \colhead{s} & \colhead{s} & \colhead{}  & \colhead{at 100\,keV} & \colhead{keV} & \colhead{} & \colhead{at 10\,MeV} & \colhead{511~keV flux\tablenotemark{c}} & \colhead{NUCLEAR flux\tablenotemark{c}} & \colhead{2.2~MeV flux\tablenotemark{c}} & \colhead{} &  \colhead{} }
      \startdata
        0 & 11:56:03.6 & 11:57:17.1 & 3.78$_{-0.03}^{+0.03}$ & 2.11$_{-0.09}^{+0.10}$ & 278$_{-23}^{+27}$ & 3.04$_{-0.11}^{+0.10}$ & 8.2$_{-0.3}^{+0.4}$ & 0.05$_{-0.04}^{+0.05}$ & 1.0$_{-0.2}^{+0.2}$ & 0.22$_{-0.06}^{+0.06}$ & 56.3/78 & 9.7e-01 \\
        \hline
        1 & 11:56:03.6 & 11:56:11.5 & 3.56$_{-0.05}^{+0.04}$ & 2.10$_{-0.13}^{+0.14}$ & 449$_{-131}^{+45}$ & 2.95$_{-0.27}^{+0.43}$ & 10.6$_{-0.6}^{+0.7}$ & 0.09$_{-0.08}^{+0.11}$ & 1.1$_{-0.4}^{+0.4}$ & 0.10$_{-0.09}^{+0.13}$ & 62.2/78 & 9.0e-01 \\
        2 & 11:56:11.5 & 11:56:19.7 & 3.57$_{-0.08}^{+0.07}$ & 2.16$_{-0.20}^{+0.25}$ & 330$_{-92}^{+148}$ & 3.07$_{-0.32}^{+0.25}$ & 8.7$_{-0.8}^{+1.0}$ & 0.08$_{-0.06}^{+0.13}$ & 0.8$_{-0.3}^{+0.4}$ & 0.22$_{-0.12}^{+0.13}$ & 75.9/78 & 5.5e-01 \\
        3 & 11:56:19.7 & 11:56:27.9 & 3.55$_{-0.09}^{+0.07}$ & 2.58$_{-0.24}^{+0.31}$ & 308$_{-74}^{+109}$ & 3.07$_{-0.20}^{+0.19}$ & 10.9$_{-1.0}^{+1.3}$ & 0.07$_{-0.06}^{+0.13}$ & 1.0$_{-0.4}^{+0.3}$ & 0.16$_{-0.11}^{+0.14}$ & 70.9/78 & 7.0e-01 \\
        4 & 11:56:27.9 & 11:56:36.1 & 3.42$_{-0.08}^{+0.05}$ & 3.92$_{-0.26}^{+0.41}$ & 276$_{-51}^{+136}$ & 3.18$_{-0.14}^{+0.10}$ & 13.3$_{-0.9}^{+1.4}$ & 0.13$_{-0.10}^{+0.16}$ & 1.4$_{-0.4}^{+0.4}$ & 0.25$_{-0.14}^{+0.12}$ & 64.0/78 & 8.7e-01 \\
        5 & 11:56:36.1 & 11:56:44.3 & 3.84$_{-0.11}^{+0.08}$ & 2.63$_{-0.29}^{+0.44}$ & 254$_{-35}^{+55}$ & 2.98$_{-0.19}^{+0.17}$ & 12.8$_{-1.4}^{+2.2}$ & 0.05$_{-0.04}^{+0.14}$ & 1.2$_{-0.3}^{+0.4}$ & 0.35$_{-0.14}^{+0.13}$ & 77.8/78 & 4.9e-01 \\
        6 & 11:56:44.3 & 11:56:52.5 & 4.22$_{-0.13}^{+0.12}$ & 1.42$_{-0.22}^{+0.28}$ & 275$_{-31}^{+43}$ & 2.50$_{-0.28}^{+0.31}$ & 25$_{-4}^{+5}$ & 0.12\tablenotemark{d} & 0.4$_{-0.3}^{+0.4}$ & 0.38$_{-0.14}^{+0.12}$ & 52.2/78 & 9.9e-01 \\
        7 & 11:56:52.5 & 11:57:00.7 & 4.57$_{-0.14}^{+0.16}$ & 1.06$_{-0.22}^{+0.25}$ & 257$_{-30}^{+34}$ & 2.34$_{-0.28}^{+0.42}$ & 27$_{-6}^{+6}$ & 0.02$_{-0.01}^{+0.11}$ & 0.4$_{-0.3}^{+0.5}$ & 0.20$_{-0.12}^{+0.12}$ & 60.3/78 & 9.3e-01 \\
        8 & 11:57:00.7 & 11:57:08.9 & 4.95$_{-0.18}^{+0.10}$ & 1.64$_{-0.23}^{+0.49}$ & 243$_{-26}^{+126}$ & 3.32$_{-1.14}^{+0.43}$ & 0.88$_{-0.12}^{+0.26}$ & 0.07$_{-0.06}^{+0.10}$ & 0.7$_{-0.4}^{+0.3}$ & 0.20$_{-0.13}^{+0.11}$ & 73.7/78 & 6.2e-01 \\
        9 & 11:57:08.9 & 11:57:17.1 & 4.66$_{-0.13}^{+0.12}$ & 1.65$_{-0.28}^{+0.36}$ & 308$_{-63}^{+46}$ & 2.6$_{-0.5}^{+0.7}$ & 9.7$_{-1.6}^{+2.1}$ & 0.06$_{-0.05}^{+0.10}$ & 0.6$_{-0.4}^{+0.3}$ & 0.11$_{-0.08}^{+0.13}$ & 67.6/78 & 7.9e-01 \\
      \enddata
    \tablenotetext{a}{In units of photons\,cm$^{-2}$\,s$^{-1}$\,keV$^{-1}$ at 100\,keV.}
    \tablenotetext{b}{In units of 10$^{-7}$\,photons\,cm$^{-2}$\,s$^{-1}$\,keV$^{-1}$ at 10\,MeV.}
    \tablenotetext{c}{In units of photons~cm$^{-2}$~s$^{-1}$.}
    \tablenotetext{d}{Upper limits.}
\end{deluxetable}
\clearpage
\end{turnpage}

\subsection{\label{sec_delays} Time Delays between Different Components}

Due to a small number of available  multichannel spectra and its coarse temporal resolution of $\sim$8\,s, the direct lag calculation using cross-correlation function is not feasible.
Instead, we used an indirect method to search for lags between parameters.
We designated the background subtracted G2 light curve available with 256-ms cadence over a reasonably long duration (longer than the time range covered by our spectral data) to be a reference light curve.
Then, we searched for a lag between this reference light curve and each distinct component in the multichannel spectra.
Specifically, for each time lags between $\sim-40$\,s and $\sim$40\,s, the reference lightcurve was rebinned to the spectra accumulation intervals, and the Pearson correlation coefficient and the corresponding $p$-value for testing non-correlation were calculated between the reference ligtcurve and each spectral parameter. The delays between each pair of different spectral components were then computed as differences between each of these components and the reference light curve.

This analysis shows that the flux in the nuclear deexcitation lines (NUCLEAR template) and the BPL amplitude at 100\,keV correlate with the G2 lightcurve at zero lag; thus, they are not delayed relative to each other. Other components show significant delays.
The amplitude of the BPL at 10\,MeV is delayed by $\sim$18\,s relative to the G2 light curve, and, accordingly, it is delayed by $\sim$18\,s relative to the NUCLEAR flux and to the amplitude at 100\,keV.
The flux of the e$^+$--e$^-$ annihilation line does not show any significant delay relative to the reference light curve; however, high uncertainties make this conclusion unreliable.
Finally, the neutron capture line is $\sim$11\,s behind the reference light curve.

\subsection{\label{sec_prot_s_neut} Proton power-law spectral index and the neutron production}


Neutrons, which penetrate in the solar atmosphere are involved in nuclear reactions.
As mentioned above, one of the most prominent reaction is the capture of a thermal neutron by a proton, p+n=$^2$H+$\gamma$, giving $E_{\gamma}$=2.223\,MeV line.
As neutrons need some time to slow down to thermal energies, 2.223\,MeV line is delayed relative to the nuclear deexcitation lines.

The time profile of the neutron capture line flux in the assumption of a weak proton spectral evolution was described by \cite{Prince1983}:

\begin{equation} \label{eq_neut_int}
F_{2.2}(t)\propto\int_{-\infty}^{t} S\left(t'\right) R(t,t') dt'
\end{equation}
where $S(t')$ is the neutron production time history, for which the prompt nuclear deexcitation line flux in 4--7\,MeV range, $F_{4-7}$, can be taken as a proxy, $R(t, t')$ is the response function, giving the 2.223\,MeV line at the time $t$ from a neutron, produced at the time $t'$. This response function can be approximated by an exponent $\exp(-(t-t')/\tau)$ with $\tau$$\sim$100\,s \citep{Murphy2007}.

\subsubsection{\label{sec_prot_s}  Time-averaged power-law index of the protons}


Most solar neutrons are born in p~--~p reactions with energy threshold $E_{thres}$$\sim$300\,MeV\,nucleon$^{-1}$ and in p~--~$^4$He, $\alpha$~--~$^4$He reactions with thresholds $E_{thres}$$\sim$30\,MeV\,nucleon$^{-1}$ and $E_{thres}$$\sim$10\,MeV\,nucleon$^{-1}$ correspondingly.
At ion energies $\sim$10\,MeV\,nucleon$^{-1}$ reactions on heavier nuclei, e. g. C, N and O, come into effect with proton ($E_{thres}$$\sim$5\,MeV\,nucleon$^{-1}$) or $\alpha$-particle ($E_{thres}$$\sim$1\,MeV\,nucleon$^{-1}$) as a projectile, but the neutron yield from these reactions is relatively low due to lower concentration of heavier nuclei \citep{Hua2002}.

Therefore, reactions involving  ions with $\sim$10--300\,MeV\,nucleon$^{-1}$ are mostly responsible for the neutron production, while the nuclear deexcitation lines in the 4--7\,MeV range are produced in interactions of $\sim$1--20\,MeV\,nucleon$^{-1}$ ions \citep{Kozlovsky2002}. Thus the ratio between nuclear deexcitation line flux and neutron capture line flux allows estimating the power-law index $s$ of the proton spectrum.

As the neutron capture line suffers from attenuation in solar atmosphere, \cite{HuaLingenfelter1987} calculated the dependence between $F_{2.2}$/$F_{4-7}$ and the flare heliocentric angle  for different proton power-law indices.
Using Figure~14 from \cite{HuaLingenfelter1987} for downward incident angular distribution of the precipitating protons and the heliocentric angle of 38$^{\circ}$ defined by the flare location,  we obtained the proton power-law index $s$=4.01$_{-0.19}^{+0.21}$; the uncertainties were estimated using Monte-Carlo modeling of the power-law indices for gaussian distributions of $F_{2.2}$ and $F_{4-7}$.

Due to the delay of 2.223\,MeV line relative to the nuclear deexcitation lines and their different time histories, the $F_{2.2}$/$F_{4-7}$ ratio can only be used for the time-averaged power-law index estimation at rather long timescales, i.\,e., at timescales longer than $\tau$$\sim$100\,s (see Eq.~\ref{eq_neut_int}).
In our case we only have $\sim$72\,s of multichannel gamma-ray observations, but according to the light curve in the hard G3 channel, \kw\ captured most of the main phase of the flare gamma-ray emission; thus, it is unlikely that the actual proton power-law index differs much from our estimation.

\subsubsection{\label{sec_neut} Time history of the neutron capture line}

This event shows an exceptionally short time delay between the flux in the nuclear deexcitation lines and the flux in the neutron capture line at 2.223\,MeV, which is usually about $\sim$100\,s.
To evaluate the expected 2.223\,MeV line temporal evolution given the flux in the nuclear deexcitation lines, we followed \cite{Kurt2017} and replaced the integral in Eq.~(\ref{eq_neut_int}) by the sum:

\begin{equation} \label{eq_neut_sum}
F_{2.2}(t_i)\propto\sum_{j=0}^{i} F_{4-7}(t_j)\exp\left(-\frac{t_i-t_j}{\tau}\right)\Delta t_j
\end{equation}

\begin{figure}\centering
\includegraphics[scale=0.57]{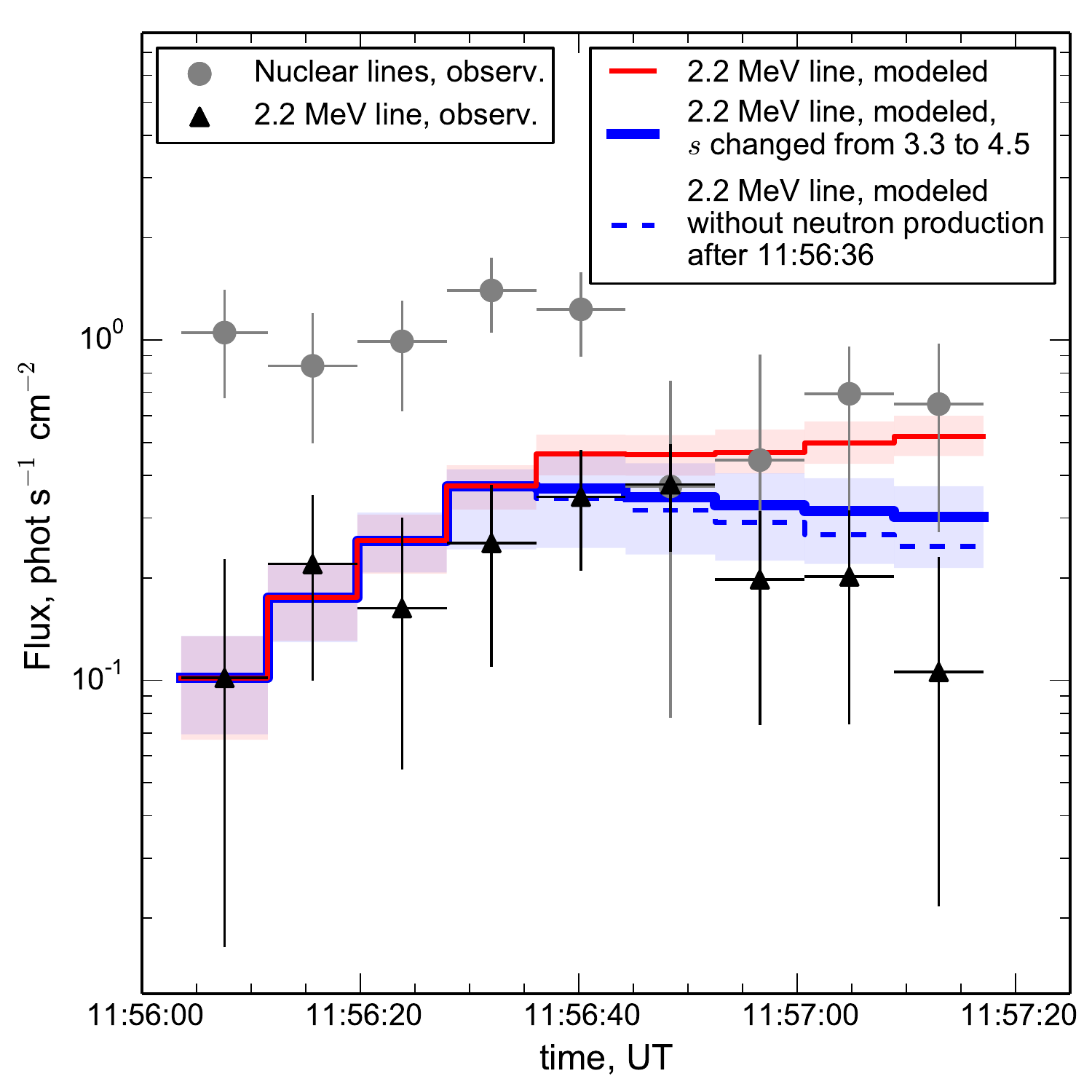}
\caption{\label{fig_neut} The observed flux in the nuclear deexcitation lines (grey circles), the observed flux in the neutron capture line 2.2\,MeV (black triangles); the modeled flux in the neutron capture line 2.2\,MeV under the assumption of a constant proton power-law index (thin red line) and its confidence bands (pink area); the modeled flux in the neutron capture line 2.2\,MeV under the assumption that the proton power-law index increases from 3.3 to 4.5 at  11:56:36 (thick blue line) and its confidence bands (light blue area); the modeled flux in the neutron capture line 2.2\,MeV without neutron production after 11:56:36 (blue dashed line). }
\end{figure}

The modeling results are presented in Figure~\ref{fig_neut}, red curve. The observed and modeled 2.223\,MeV line fluxes were normalized for the first time interval. The confidence intervals for modeled flux (pink area) were estimated using Monte-Carlo modeling for gaussian uncertainties in the NUCLEAR flux.
During the first six time intervals, the observed and modeled fluxes are in good agreement with each other, but after 11:56:52 the observed values undergo a sharp decrease. This can be explained by an abrupt reduction of the neutron yield.

\begin{figure}\centering
\includegraphics[scale=0.57]{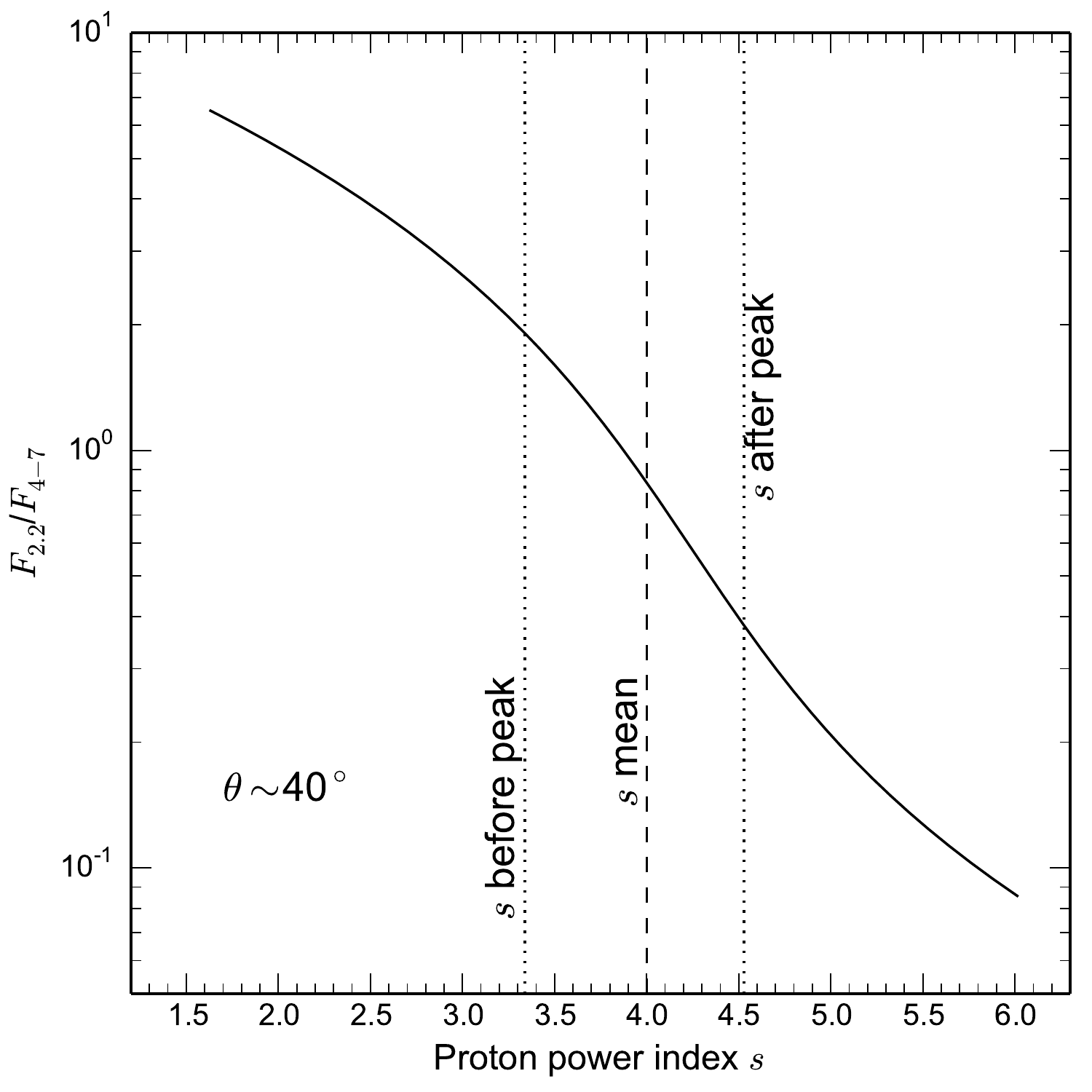}
\caption{\label{fig_f22f47_s} Dependence between the ratio $F_{2.2}$/$F_{4-7}$ of the proton power-law index $s$ for the heliocentric angle $\theta$$\sim$40$^{\circ}$. The dashed line indicates the flare-averaged power-law index, the dotted lines correspond to the upper limit for power-law index before the peak and lower limit after the peak. 
}
\end{figure}

As follows from Section~\ref{sec_prot_s}, the neutron production rate strongly depends on the spectral slope of the accelerated ion spectrum and the  abundances \citep{Murphy2007}.
The dependence between the $F_{2.2}$/$F_{4-7}$ ratio and the proton power-law index is presented in Figure~\ref{fig_f22f47_s}.
This curve is obtained by the 3rd order polynomial fitting of the digitized data on Figure~14 in \cite{HuaLingenfelter1987} for the downward incident angle distribution and the heliocentric angle $\theta$$\sim$40$^{\circ}$.
As the 2.223\,MeV line yield goes down as the spectral index increases, the probable reason for the neutron number reduction and, thus, for the rapid decline of the 2.223\,MeV line flux, could be a sudden steepening of the nonthermal ion spectrum.
We model the neutron capture line time profile in assumption, that the neutron production was reduced sharply by $r_{n}$ times at one of the following moments $t_n$ during the flare: the end of interval \#~3, the middle of interval \#~4, the end of interval \#~4, the middle of interval \#~5 or the end of interval \#~5.
The best model-to-data agreement is obtained for the $t_n$ taken as the end of the interval \#~4 and for the neutron yield reduction by at least $r_n$=5 times. The associated modeled flux is plotted in Figure~\ref{fig_neut}, blue curve.
The confidence intervals (light blue area) are obtained using Monte-Carlo modeling of the flux in the $NUCLEAR$ template within an assumption of its gaussian distribution and $t_n$ uniformly distributed during time intervals \#~4 and \#~5.
For $r_n$>10 the 2.223\,MeV line profile becomes insensitive to the further increase of $r_{n}$  and reaches its limit, where no neutrons are produced after 11:56:36 (dashed blue line in the Figure).

Based on the decrease of the 2.223\,MeV line flux by $r_{n}$>5 times, the time-averaged proton power-law index $s\sim$4, and using Figure~\ref{fig_f22f47_s} as a guide, we obtain an upper limit of the proton power-law index before the steepening and its lower limit after the steepening, which turned out to be $\sim$3.3 and $\sim$4.5, correspondingly.
These limits for the proton power-law indexes are shown in Figure~\ref{fig_mult}(b).

Alternatively, the implied decrease of the neutron yield can be explained by a sudden change of the accelerated particle abundances (see Section~\ref{sec_prot_s}), if, for example, the ratio of accelerated $\alpha$-particles to protons $\alpha$/p decreased from 0.5 before the the flare peak to 0.1 after the peak, which can result in $r_{n}$$\sim$5 \citep{Murphy2007}; however, such a dramatic change in the abundances of the accelerated particles at the course of the flare looks unlikely.

\subsection{\label{sec_ion_en} Ion energetics}

Accelerated ions may significantly contribute to the total flare energetics \citep{Emslie2012}.
Let us estimate the energy released by the accelerated protons during the \kw\ observational time in the gamma-ray range.

According to the \verb"OSPEX" NUCLEAR template 
model\footnote{\url{https://hesperia.gsfc.nasa.gov/ssw/packages/spex/idl/object\_spex/fit\_model\_components.txt}}, the flux 
$F_{nuclear}$ 
of 1\,phot.\,cm$^{-2}$\,s$^{-1}$ at the Earth corresponds to the production rate of $N_p$=7.60$\times$10$^{29}$ protons s$^{-1}$ with energies $\ge$$E_{norm}$=30~MeV at the Sun in the assumption of the proton power-law spectrum with the index $s_{norm}$=4, which agrees well with what we obtained in Section~\ref{sec_prot_s}.

Under assumption that the proton spectrum is a single power law with a low-energy cutoff at $E_{cut,lo}$=1\,MeV \citep{Emslie2004, Emslie2012}, we obtain that the proton energy of $E$$\sim$3.6$\times$10$^{30}$\,erg has been released during $\sim72$~s of the flare impulsive phase captured by \kw .
Accelerated protons contain only an uncertain fraction of the energy of all accelerated ions/nuclei.
The total ion energy depends on both elemental abundances and spectral shapes of various species of the accelerated ions. Apparently, we do not have a sufficient number of observational constraints to properly estimate the total energy of the flare-accelerated ions.
Instead, to report a number that can directly be compared with previous reported estimates \citep{Emslie2012}, we assume the total ion energy to be three times larger than the proton energy; thus, we estimate the total energy of ions accelerated above 1~MeV to be $\sim$11$\times$10$^{30}$\,erg.

Our spectral data cover only a portion of the impulsive phase of the flare; thus, we estimated only a portion of the total ion energy during the impulsive phase.
Given the time history in the G2 channel, it is, however, unlikely that we underestimated this energy by a factor larger than 1.5--2.

\section{\label{sec_disc} Discussion}

Unlike countless X-ray solar flares, the total number of solar flares recorded in the gamma-ray range is still limited to a few dozens events.
For this reason, a detailed analysis of each gamma-ray solar flare is potentially highly valuable to quantify acceleration of ions in solar flares, when signatures of gamma-ray lines are present in the spectrum along with the bremsstrahlung continuum.
The question, if the gamma-ray lines are present in the spectrum or not is challenging, especially when dealing with a moderate-resolution data, which are not capable of resolving most individual lines.
In this paper, for the first time, we employed a Bayesian inference to objectively conclude about the presence of distinct spectral components in the solar gamma-ray data.
The Bayesian inference confirmed the presence of the nuclear deexcitation lines, the electron-positron annihilation line at 511\,keV, and the neutron capture line at 2.223\,MeV in the 2017-Sep-6 X9.3 flare, which is an unambiguous evidence that ions were accelerated during the impulsive phase of this event.

Relationships between different spectral components are sensitive to the spectrum of accelerated ion; in our case the flare-averaged power-law index of the accelerated protons is $\sim$4, which is consistent with an averaged value of gamma-ray flares detected by Solar Maximum Mission \citep{Ramaty1996}.
The spectral evolution of the nonthermal proton population can be nailed down from timing of various spectral components.
The neutron capture line at 2.223\,MeV is delayed relative to the nuclear deexcitation lines as expected. However, the magnitude of this delay is only $\sim$11\,s, which is an order of magnitude shorter as compared to the typical value of $\sim$100\,s.
Such a short time delay is consistent with a sharp decrease of the neutron production at the course of the flare, which might be caused by a reasonably fast softening of the proton spectrum, where the ion power-law index $s$ increases by at least $\sim$1.2 with $s<3.3$ in the beginning of the impulsive phase, while $s>4.5$ afterwards.

The spectral evolution of the bremsstrahlung continuum is also interesting.
The continuum photon component is consistent with a broken power law with a break-up at 300--500~keV during the entire duration of the impulsive phase.
During roughly first half of it, the low- and high-energy spectral slopes evolve consistently and show a modest difference between each other, $\Delta\gamma\lesssim0.5$, which is consistent with a bremsstrahlung spectrum from a single power law distribution of nonthermal electrons, if one takes into account emission at electron-electron collisions, which becomes important in the relativistic regime.
However, during the second half of the impulsive phase, the low-energy part of the spectrum softens showing a typical soft-hard-soft (SHS) pattern, while the high-energy component further hardens showing a soft-hard-harder (SHH) pattern.
Although we cannot offer a unique interpretation of this behavior, we envision two possibilities.
One of them involves a more prolonged acceleration of electrons originally accelerated up to a few hundred keV by an additional (second-step) acceleration.
The other one associates this high-energy component with a bremsstrahlung produced by secondary relativistic positrons.
These positrons, as we mentioned in the Introduction, can be produced by either $\pi^+$ decay or $\beta^+$ decay or both.
Given that those positrons are born with high, often relativistic, energy, no additional acceleration is needed (though, not excluded).
Instead, the observed spectral flattening can be ascribed to Coulomb losses, whose efficiency goes down with energy.
This second option is easier to reconcile with the observed softening of the proton spectrum, which happens at this stage.
This is also consistent with the observation that the accelerated ions and electrons both show similar SHS patterns, which is expected if both electron and ions are accelerated by the same or closely related acceleration mechanisms. Having the same acceleration mechanisms for electrons and  ions is further consistent with
the detected high correlation between the nuclear deexcitation line time profile and the bremsstrahlung continuum for zero time lag.
However, even though the contribution from relativistic positrons to bremsstrahlung continuum at higher energies offers an appealing explanation for this particular case, it can hardly be applied for the ``electron-dominated'' flares. Indeed, in the electron-dominated events there is no evidence of the gamma-ray lines, which implies no or only a weak ion acceleration; thus, the positron yield is expected to be also weak or non-existent.

The total energy of the accelerated ions has been estimated as $\sim$11$\times$10$^{30}$\,erg, which is in the range between 0.19$\times$10$^{30}$ and >190$\times$10$^{30}$ obtained by \cite{Emslie2012} for 21 flares.
As \kw\ did not observe the entire duration of the gamma-ray emission, the actual ion energetics can be larger; it is however, unlikely that we underestimated this energy by more than a factor of two.


\section{Conclusions}
The main conclusion is that ions were accelerated during the impulsive phase of the 2017-Sep-06 X9.3 flare.
Accelerated ions demonstrate a typical power-law index of 4 and ``mean'' energetics ($\sim$11$\times$10$^{30}$\,erg) as compared to other flares, although the accelerated proton population demonstrates a prominent softening at the course of the flare.
Spectral evolution of the lower energy part of the continuum is similar to that of the accelerated protons.
In contrast, the  spectral slope of the continuum at higher energies does not correlate with any of them, while becomes harder at the course of the flare.

\acknowledgements
We are highly grateful to Prof. Gerald Share and Prof. Ronald Murphy for productive discussions.
A.L.L. (\kw\ data preparation, modeling, estimation of the accelerated ion properties) gratefully acknowledges support from RFBR grant 18-32-00439.
G.D.F. was supported in part by NSF grant AGS-1817277 
and NASA grants
80NSSC18K0667, 
80NSSC19K0068, 
and 80NSSC18K1128 
to New Jersey Institute of Technology. S.A.A. (Bayesian inference, MCMC, model comparison) was supported by the RFBR grant 17-52-53203 GFEN\_a.

\appendix

\section{\label{app_pileup} Pulse pile-up effect (PPE) correction}
\begin{figure}\centering
	\includegraphics[scale=0.7]{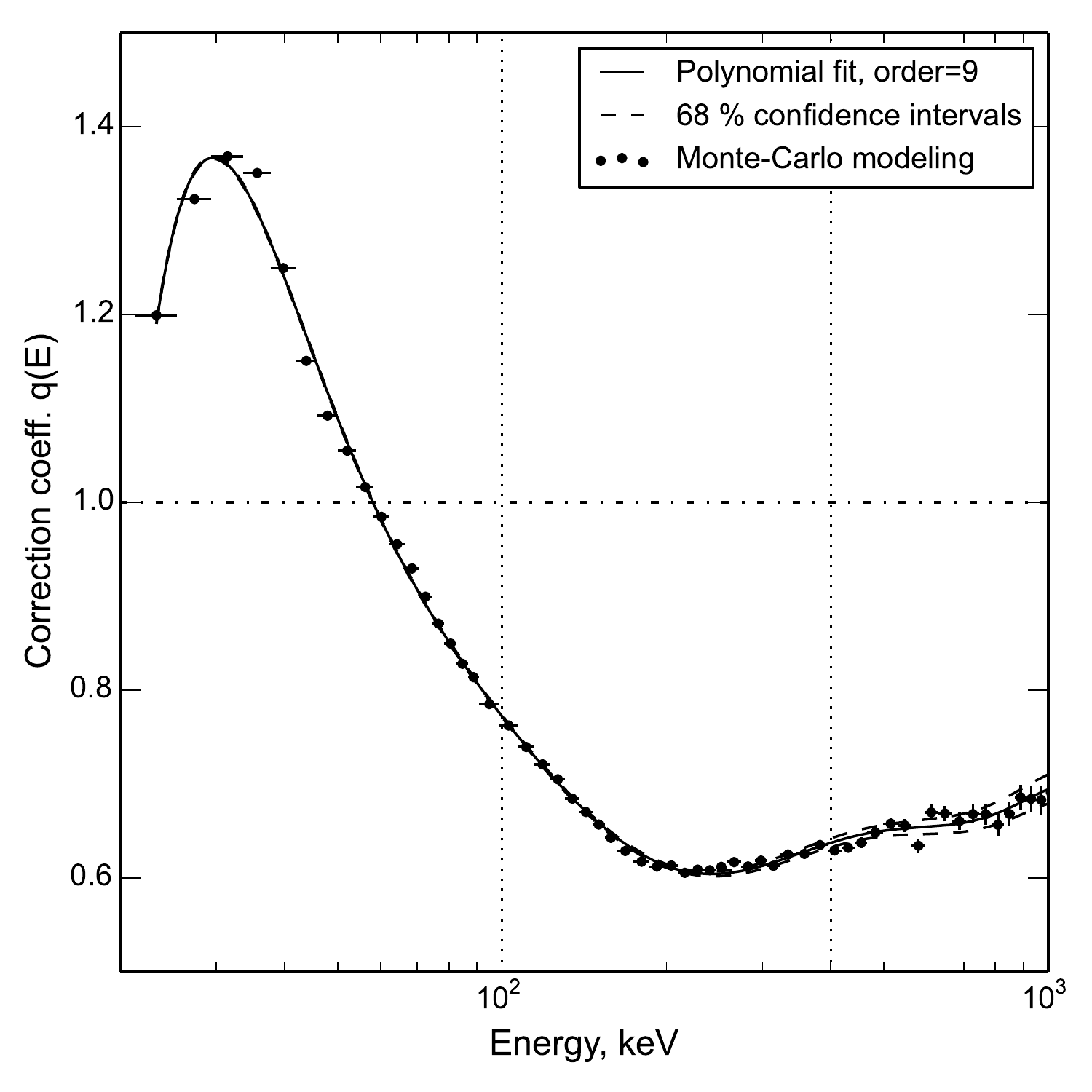}
	\caption{Pile-up corrections for spectrum \#4: simulated data points and a 9-th degree polynomial approximation. Vertical dotted lines hightlight considered energy interval 100--400\,keV, horizontal dash-dotted line marks correction coefficient=1.
		\label{fig_ppu}}
\end{figure}

The approach used in this work is based on an iterative procedure that aims on minimizing the difference between
the pileup-distorted instrumental spectrum and the simulated count spectrum.
The latter is produced, given an incident photon flux and a specific spectral shape,
by a specially developed Monte-Carlo (MC) simulation software (hereafter, {\tt\string PPU}),
which models the behavior of the instrument's analog and digital electronic circuits at high count rates.
The {\tt\string PPU} code traces its origin to the software used in an analysis of the \kw\ detection
of the 1998 August 27 giant flare from Soft Gamma-repeater SGR 1900+14 (details of these simulations
and the \kw\ dead-time correction procedures can be found in \citealt{Mazets1999}).

To obtain spectral correction parameters we performed the following steps:
\begin{enumerate}
\item An instrumental, PPE-distorted count spectrum ($\textbf{\textit{S}}_0$) was fitted in {\tt\string XSPEC}
by a suitable spectral model (BPL, in our case) and an initial set of spectral parameters ($\textbf{\textit{P}}_0$) has been obtained.
This step is necessary to estimate the spectrum hardness, on which the PPE distortion pattern is highly dependent.

\item An artificial, non-distorted count spectrum $\textbf{\textit{S}}_1$
was generated assuming the spectral parameters $\textbf{\textit{P}}_0$ using the {\tt\string fakeit} option of {\tt\string XSPEC}.

\item An artificial PPE-distorted count spectrum $\textbf{\textit{S}}_2$ was simulated by the {\tt\string PPU} routine
given the count rate in $\textbf{\textit{S}}_0$ and assuming the shape of $\textbf{\textit{S}}_1$ as a probability distribution for the incident pulse amplitudes.

\item Spectra $\textbf{\textit{S}}_1$ and $\textbf{\textit{S}}_2$ were both normalized to unity area and the pile-up correction coefficient
$q_i$ was estimated for each spectral channel $i$ as the $\textbf{\textit{S}}_{1,i}$/$\textbf{\textit{S}}_{2,i}$ ratio in the channel.
Since the pileup-caused distortion is supposed to be a smooth function of a channel energy $E$,
in further calculations, the PPE-correction factor $q(E)$ is taken as a polynomial interpolation
of $q_i(E_i)$, thus evening out fluctuations resulting from the MC simulations.

\item As the result, a PPE-corrected count spectrum $\textbf{\textit{S}}_c$ is calculated as $\textbf{\textit{S}}_c(E_i) = \textbf{\textit{S}}_0(E_i)*q(E_i)$.
\end{enumerate}

The major weakness of the described procedure lies in the estimation of the spectral hardness from the PPE-distorted spectrum $\textbf{\textit{S}}_0$.
An importance of this issue has been checked by adding a second iteration,
i.e. by using spectral parameters $\textbf{\textit{P}}_c$ estimated from the PPE-corrected spectrum $\textbf{\textit{S}}_c$
as an input to the additional round of the simulations: $\textbf{\textit{S}}_c \rightarrow \textbf{\textit{P}}_c \rightarrow \textbf{\textit{S}}_{1,c} \rightarrow \textbf{\textit{S}}_{2,c} \rightarrow q\prime(E)$.
We found the resulting second order approximation $q\prime(E)$ not much different from the first order $q(E)$ and used the latter in practical calculations.

We tested the described approach on distorted spectra, simulated in {\tt\string PPU} for a number of BPL parameter sets,
and found, that, at incident count rates up to ~$2 \times 10^5$ cts/s, our correction procedure allows to recover
BPL-like spectral shapes to the accuracy of $\lesssim$0.1 in the low- and high-energy spectral indices, and $\sim$10~keV in the break energy.

For the most distorted spectrum of the 2017-Sep-06 X9.3 flare (\#4), at the incident count rate of ~$1.8 \times 10^5$ cts/s,
the scale of the PPE correction $q$ varies with $E$ from $\sim$1.4 at $\sim$30~keV to $\sim$0.6 at $\sim$200~keV (Fig.~\ref{fig_ppu}),
assuming a BPL with $\gamma_1 \sim 2.5$, $\gamma_2 \sim 3.3$, and $E_\mathrm{break} \sim120$ keV as the incident spectrum.

In the low-energy channels, the thermal bremsstrahlung can play a significant role, but \kw\ cannot recover
the thermal plasma parameters due to very few channels and instrumental glitches in this range.
In most cases GOES X-ray sensors in two broadband SXR channels, 1–8\,\AA\ and 0.5–4.0\,\AA\ can be used \citep{White2005}, but for intense hot flares both electron temperature and emission measure are likely underestimated \citep{Caspi2014}.
To estimate thermal plasma parameters we used relationships from Table 1 in \cite{Caspi2014} between
GOES class and temparature and GOES class and emission measure for hot flares.
Regression for an X9 class flare gave the temperature $\sim$48\,MK and emission measure $\sim$5$\times$10$^{48}$\,cm$^{-3}$.
We added to the artificial spectra S1 thermal bremsstrahlung component with these parameters, but the appropriate PPE-correction factor $q(E)$ did not changed significantly in the considered energy interval from 100 to 400\,keV, thus we decided to stay on BPL model for obtaining the correction coefficients.



\clearpage
\bibliography{X9_2017Sep6}

\end{document}